\documentclass[twocolumn,showpacs,floats,floatfix,superscriptaddress,aps,pra]{revtex4}

\usepackage{amsfonts}
\usepackage{amssymb}
\usepackage{amsmath}
\usepackage{graphicx}
\usepackage{bm}

\begin{document}

\author{Elica Kyoseva}
\affiliation{Centre for Quantum Technologies, National University of Singapore, 2 Science Drive 3, Singapore 117542}
\author{Dimitris Angelakis}
\affiliation{Centre for Quantum Technologies, National University of Singapore, 2 Science Drive 3, Singapore 117542}
\affiliation{Science Department, Technical University of Crete, Chania, Crete, Greece, 73100}
\author{Kwek Leong Chuan}
\affiliation{Centre for Quantum Technologies, National University of Singapore, 2 Science Drive 3, Singapore 117542}
\title{A single-interaction step implementation of a quantum search \\
in coupled micro-cavities}
\date{\today}


\begin{abstract}
We present a method for realizing efficiently Grover's search algorithm in an array of coupled cavities doped with three-level atoms. We show that by encoding information in the lowest two ground states of the dopants and through the application of appropriately tuned global laser fields, the reflection operator needed for the quantum search algorithm can be realized in a single physical operation. Thus, the time steps in which Grover's search can be implemented become equal to the mathematical steps $\sim \mathcal{O}(\sqrt{N})$, where $N$ is the size of the register. We study the robustness of the implementation against errors due to photon loss and fluctuations in the cavity frequencies and atom-photon coupling constants.
\end{abstract}

\maketitle

The potential to execute certain types of algorithms much more efficiently than the corresponding classical counterparts is one major reason why quantum computers were initially proposed. To date, several important quantum algorithms have been invented: the two most well known ones being the quantum search and the prime number factorization \cite{algorithms}. To realize a physical implementation of a quantum algorithm, it is important that one can encode the basic units of quantum information, i.e the qubits, initialize them to some suitable inputs, perform an adequate set of unitary operations and then finally read the output. Many physical implementations have been proposed and performed based on trapped ions, cold atoms and solid-state systems \cite{books} with varying degree of success for realizing few-qubit applications. More recently, coupled cavity arrays have been proposed as a new hybrid light-matter system for quantum simulation applications. This system was initially studied for conditional photon phase gates \cite{angelakis04} and later for Mott transitions \cite{angelakis07,hartmann06,greentree06}. Following the initial works, a number of papers appeared studying the details of the polaritonic many-body state and simulations of more complex spin models were proposed \cite{hartmann07,rossini07,paternostro07,cho07,li08,kay08,Chang,Tanabe,Aichhorn,Ji}. More recently driven arrays were considered towards the production and coherent control of steady state entanglement \cite{two-state} under realistic dissipation parameters. Also, an analogy with Josephson oscillations was shown and the many body properties of the driven array have been recently studied \cite{coherent-control-of-photon-emission}.

In the current work we present for the first time a method for realizing Grover's search algorithm for an array of coupled cavities in an \emph{efficient} way. We start by noting that the basic Grover iteration is executed in two steps -- first, the oracle $\mathrm{O}$ is applied to mark the searched state by flipping its phase, and then a global reflection operation about the mean $\mathcal{G}(W)= \mathrm{1} - 2|W\rangle \langle W|$ is performed, where $W$ is the $N$-qubit $W$ state. We will show that by exploiting the natural evolution of our system we can implement the reflection operation $\mathcal{G}(W)$ -- also known as a quantum mirror or Householder reflection \cite{Householder} -- in only \emph{one physical step}. The quantum reflection, together with the oracle operation $\mathrm{O}$, comprise the Grover logical step which is performed $\sim \mathcal{O}(\frac{\pi\sqrt{N}}{4})$ times to find the searched entry. The current techniques for performing these transformations utilize sequences of single- and two- qubit gates in order to perform an $N$-qubit quantum gate \cite{Zeilinger}. Hence, the number of operations scale as $\sim \mathcal{O}(N^{2})$ with the size of the register. Here, we will exemplify how to perform the required $N$-qubit reflection gate for the Grover iteration in only one operational step. In this sense the required physical time steps are significantly reduced and become equivalent to the number of mathematical steps $\sim \mathcal{O}(\frac{\pi \sqrt{N}}{4})$.

\begin{figure}[t]
\centering
\includegraphics[width=85mm]{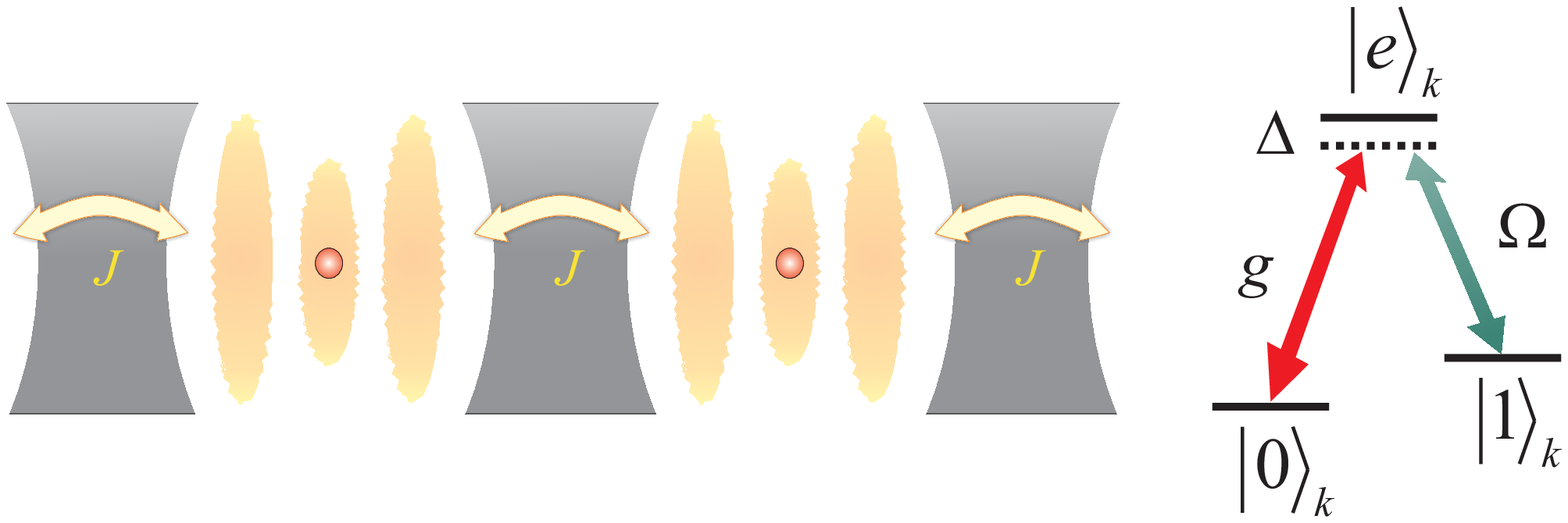}
\caption{An array of coupled cavities via hopping photons between neighbouring sites at a rate $J$ doped with three-level atomic lambda systems.}
\label{cavities}
\end{figure}

Consider a linear chain of $N$ cavities fabricated in such a way that the spatial profile of their cavity modes overlap and photons can hop between neighbouring sites. Let each cavity be doped with a three-level atomic system with a lambda internal energy configuration, as shown in Fig. 1 and let us assume that the $N$ logical qubits, which we utilize for the quantum register, are encoded in the ground states $|0\rangle_{k}$ and $|1\rangle_{k}$ of the individual atoms \footnote{We use the term atoms but this can applied to a variety of technologies including superconducting qubits, ions or quantum dots.}. The $|0\rangle_{k} \leftrightarrow |e\rangle_{k}$ transition of each atom is coupled to the respective cavity mode with coupling strength $g$ and detuning $\Delta=\omega_{0}-\omega_{c}$ being the difference between the Bohr transition frequency of the atom $\omega_{0}$ and the cavity frequency $\omega_{c}$. Additional laser fields drive the $|1\rangle_{k} \leftrightarrow |e\rangle_{k}$ transitions with Rabi frequency $\Omega$  and are detuned from the transition frequency again by $\Delta$. The total Hamiltonian which governs the evolution of the system consists of three parts -- a free energy term, an interaction term describing the coupling between the atoms and the respective cavity modes, and a hopping term. Using the dipole and the rotating wave approximations we can express the system Hamiltonian in an appropriately chosen interaction picture as
\begin{eqnarray}
\mathsf{H}^{\textnormal{total}} &=& \sum_{k=1}^{N}\omega_{c} a^{\dag}_{k}a_{k} + J \sum_{k=1}^{N}(a^{\dag}_{k}a_{k+1} + \mathrm{H.c.}) \notag \\
&+& g\sum_{k=1}^{N} (\mathrm{e}^{-\mathrm{i}\Delta t} |0\rangle_{k}\langle e|_{k} a_{k}^{\dag} + \mathrm{H.c.}) \notag \\
&+& \frac{1}{2}\Omega \sum_{k=1}^{N} (\mathrm{e}^{-\mathrm{i}\Delta t} |1\rangle_{k}\langle e|_{k} + \mathrm{H.c.}),
\label{Ham1}
\end{eqnarray}
where we have adopted the convention $\hbar=1$. Without loss of generality we can assume that $\Omega$ and $g$ are real, by including their phases in the atomic and cavity photon states.

For the one step realization of the Grover search algorithm we will exploit the single-excitation subspace and initialize all qubits in their ground states $|0\rangle_{k}$ while populating the lowest energy common photonic mode with one photon. Hence, the available Hilbert subspace is $(N+1)$-dimensional and is spanned by the states $|\psi_{n};0\rangle$, $(n=1,...,N)$ and $|\psi_{0};1\rangle$. Here, $|\psi_{n};0\rangle$ labels a collective atomic state in which the $n^{\textnormal{th}}$ atom is in state $|1\rangle$, all other atoms are in states $|0\rangle$, and the mode has 0 photons; while $|\psi_{0};1\rangle$ corresponds to all qubits in states $|0\rangle$ and one photon populating the common mode.

To implement the required operations we exploit the direct coupling of the logical qubits via the common cavity photon modes, which we denote as $A^{\dag}_{j} \; (A_{j})$ and are obtained through a Fourier transform of the local modes. Our aim is to achieve a relatively strong coupling between the logical qubits via the common cavity modes. We therefore consider a parameter regime in which the excited atomic states $|e\rangle_{k} \; (k=1,...,N)$ evolve on a much faster time scale than all other states, $J \ll \Omega, \;g  \ll \Delta$. This allows us to adiabatically eliminate these states from the dynamics of the system. In this case the Hamiltonian (\ref{Ham1}) for qubits coupled to a single common mode $A_{1}^{\dag}$ simplifies to
\begin{equation}
\mathsf{H}^{\textnormal{eff}}=\sum_{k=1}^{N} g^{\prime} |0\rangle_{k} \langle1|_{k} A^{\dag}_{1} \mathrm{e}^{-\mathrm{i} \Delta^{\prime} t} + \mathrm{H.c.}
+ \delta A^{\dag}_{1}A_{1} |0\rangle_{k} \langle0|_{k}
\label{Ham2}
\end{equation}
with effective coupling constants $g^{\prime} =-g \Omega/2\Delta$, and detuning $\Delta^{\prime}=-\Omega^{2}/4\Delta$. Moreover, $\delta =-g^{2}/(\Delta - 2J)$ is the detuning of the coupling of the qubits to the lowest energy Bloch mode $A_{1}^{\dag}$.

Performing a simple phase transformation the Hamiltonian (\ref{Ham2}) can be rewritten in terms of the collective atomic states as
\begin{equation}
\mathsf{H}^{\textnormal{eff}} = \sum_{k=1}^{N}g^{\prime}|\psi_{k};0\rangle \langle \psi_{0};1| + \mathrm{H.c.} + (\Delta^{\prime}-\delta) |\psi_{0};1\rangle \langle \psi_{0};1|.
\label{Ham3}
\end{equation}
Note that the above Hamiltonian describes the interaction between $N$ degenerate states, corresponding to $|\psi_{k};0\rangle$, coupled to one excited state $|\psi_{0};1\rangle$ by coupling constants $g^{\prime}$ and detuned by $(\Delta^{\prime}-\delta)$. We assume that the qubits are addressed by a global laser pulse which automatically fulfills the requirements for equal couplings and detunings. For a system governed by the Hamiltonian from Eq. (\ref{Ham3}) Ref. \cite{Kyoseva06} provides an analytical expression for the propagator $\mathsf{U}(t,0)$ which determines the system's state at a moment $t$ according to $|\Psi(t)\rangle= \mathsf{U}(t,0)|\Psi(0)\rangle$. In the basis of the states $\{|\psi_{n};0\rangle,|\psi_{0};1\rangle\}$ $\mathsf{U}(t,0)$ is given by
\begin{eqnarray}
\mathsf{U}(t,0)&=&\sum_{i\neq j=1}^{N} \{ \delta_{ij} + \big( a(t)-1 \big) \frac{g^{\prime2}}{\chi^{2}} \} |\psi_{i};0\rangle \langle\psi_{j};0| \notag \\
&+& a^{*}(t) |\psi_{0};1\rangle \langle\psi_{0};1| \notag \\
&+& b(t)\sum_{i=1}^{N}\frac{g^{\prime}}{\chi} |\psi_{i};0\rangle \langle\psi_{0};1| - \mathrm{H.c.}
\label{U}
\end{eqnarray}
Here we have introduced the collective atomic Rabi frequency $\chi = \sqrt{N}g^{\prime}$. The complex parameters $a$ and $b$ depend on the interaction between the qubits and the applied external fields. They can be found analytically or numerically for any set of detunings, pulses' shapes and amplitudes $\{\Delta^{\prime}-\delta,g^{\prime}\}$; and obey the relation $|a|^{2}+|b|^{2}=1$. Some values of interest of $a$ and $b$ are analyzed in Ref. \cite{Kyoseva06}.

Now, we proceed with the execution of the Grover algorithm. First, we describe the \emph{initialization} of the qubits in a $|W\rangle = \sum_{n=1}^{N}|\psi_{n}\rangle /\sqrt{N}= [1,\ldots,1]^{\textnormal{T}}/\sqrt{N}$ state, which corresponds to an equally weighted superposition of all states. It can be created in a \emph{single physical step} just through the natural time evolution of Eq. (\ref{U}) starting from $|\psi_{0};1\rangle$ and applying a global laser pulse with a hyperbolic-secant time dependence, a pulse width $T$ and an area of $\pi$, while satisfying $\Delta^{\prime}-\delta=0$. Hence, the typical width of this pulse is $T = \frac{\pi}{\chi} =\frac{\pi}{\sqrt{N} g^{\prime}} \simeq 10^{-7} s$, for the estimation of which we have used an experimentally observed individual coupling strength $g\simeq 105$ MHz from Ref. \cite{112} and we have assumed that $\Omega \sim g$ and $\Delta \sim 10 g$. For this interaction step the parameter $a=0$ and the state of the system thereafter is given by $|\Psi\rangle= \sum_{i=1}^{N} g^{\prime}|\psi_{i};0\rangle/\chi$.

\begin{figure}[t]
\center
\includegraphics[width=85mm]{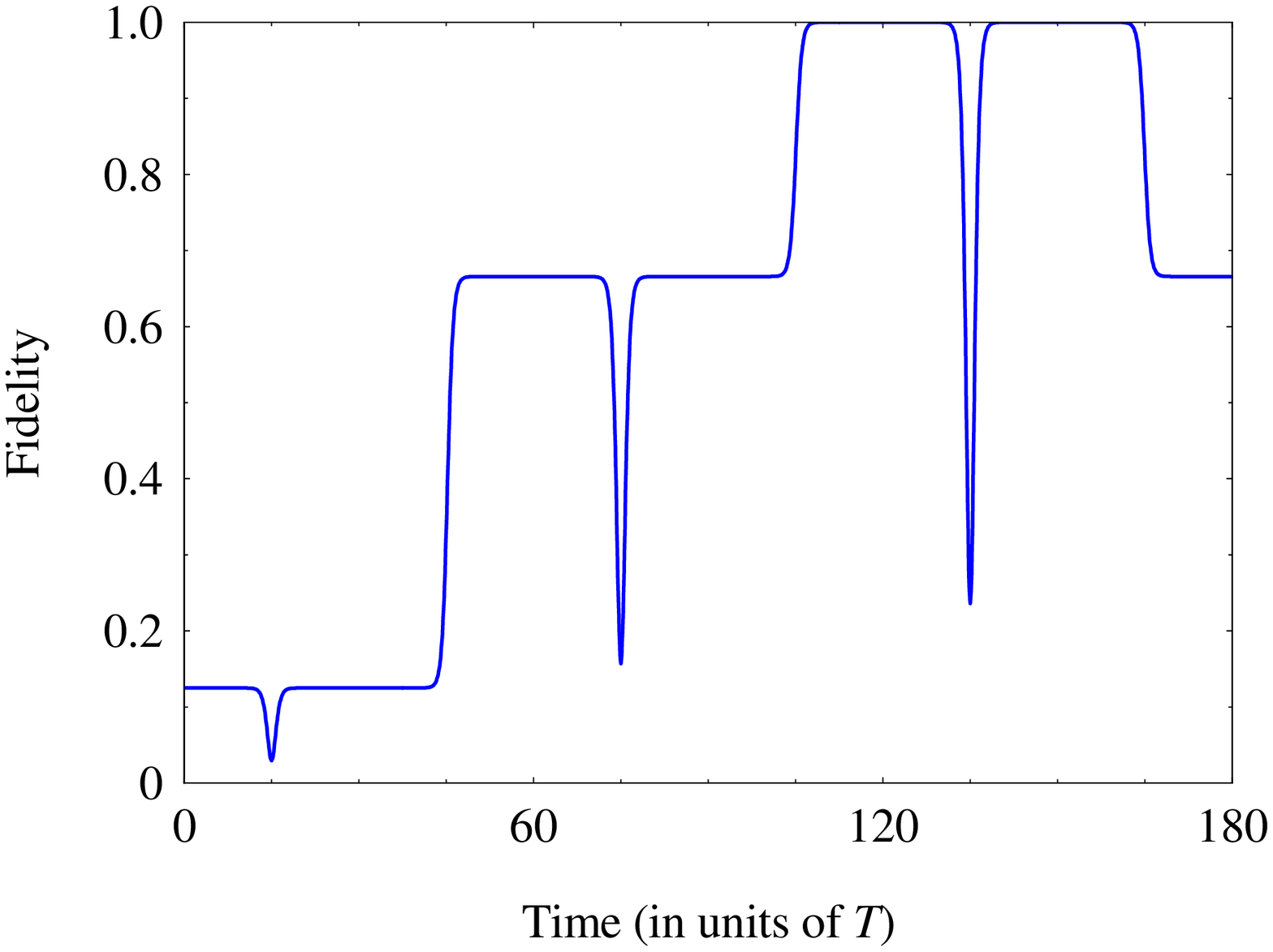}
\caption{Numerical simulation of the population of the searched state for the probabilistic Grover search implemented in an array consisting of $N=8$ coupled cavities. The system is initially prepared in an eight-qubit $W$ state through the application of a single global pulse with an area of $\pi$. After the application of two Grover iterations each comprising an oracle operation (implemented by a \emph{local} laser pulse of area $2\pi$) and a global reflection about the mean (realized with a \emph{global} pulse of area $2\pi$ and ) the fidelity of the state approaches unity. After that it starts to decrease and shows oscillatory behaviour. The sharp disturbances of the plotted curve at $15T$, $75T$ and $135T$ indicate the application of the oracle operations while the distinct increases (decreases) at $45T$, $105T$ and $165T$ show the global laser pulses which were chosen to have sech time dependence and a pulse area of 2$\pi$.}
\label{fig2}
\end{figure}

After that we continue with the description of the \emph{implementation of the Grover iteration}. Mathematically, the Grover logical operation consists of two steps -- a local oracle operation $\mathrm{O}$ which marks the searched state by flipping its phase and a global reflection about the mean. Once we have prepared the required $N$ qubit $|W\rangle$ state the oracle operation $\mathrm{O}$ can be easily implemented by a \emph{local} laser pulse, addressing only the searched qubit, with a pulse area of $2\pi$. The resulting state of the system is then evolved according to the propagator given in Eq. (\ref{U}) as we apply a consequent \emph{global} pulse, addressing all qubits, in order to realize the Householder reflection. It is again assumed to have a hyperbolic-secant shape and an area of 2$\pi$ (hence, the individual qubits experience pulse areas of $2\pi/\sqrt{N}$) which effectively implements a 2$\pi$ rotation between the states $\sum_{i=1}^{N}|\psi_{i};0\rangle \leftrightarrow |\psi_{0};1\rangle$. Then, the parameter $a=-1$ and the propagator for the $N$ degenerate states (\ref{U}), comprising the database of the system, becomes
\begin{equation}
\label{G}
\mathcal{G}(W)= \mathrm{1} - 2|W\rangle \langle W|,
\end{equation}
which is the $N$-dimensional Householder reflection operator about the mean \cite{Ivanov2}. Here, the normalized reflection vector $|W\rangle=\frac{1}{\sqrt{N}}[1,1,\ldots,1]^{\textnormal{T}}$ is the $N$ qubit $|W\rangle$ state. Hence, in a \emph{single physical step} we can realize in the register's subspace the basic ingredient for the Grover's algorithm just by tuning accordingly the interaction parameters of the system. Moreover, it has been shown \cite{Ivanov} that \emph{any} desired $N$-qubit unitary operation can be realized as a sequence of reflection operations (\ref{G}) in a number of steps $\sim \mathcal{O}(N)$, which is an improvement compared to methods based on, for example, sequences of two-dimensional rotations \cite{Zeilinger}. The Grover logical step $\mathcal{G}(W)\mathrm{O}$ is successive executed  $\pi\sqrt{N}/4$ times on the register which drives the system into the marked state.

The third final step in the search algorithm is the \emph{detection}. The marked qubit is the only one in state $|1\rangle$ while all others are in state $|0\rangle$. This can be easily probed by employing usual atomic state measurement techniques \cite{Rowe}. In Fig. \ref{fig2} a numerical calculation of the fidelity of the preselected state is plotted as a function of time for an array consisting of $N=8$ cavities. After the application of two Grover iterations, i.e. two oracle calls and two global reflections, almost all population is driven into the marked state, with fidelity $0.98\%$. Then, it starts to decrease as a part of oscillations between zero and one.

In an experimental setup in addition to the conditions $J \ll \Omega, g \ll \Delta$ we also need to assume that the cavity leakage time is smaller than the time required to implement the logical/physical steps. In our case this translates to $\chi < \gamma_{\textnormal{cav}}$, i. e., close to strong coupling regime which corresponds to single-cavity cooperativity parameter $C = \frac{g^{2}}{\Gamma \gamma_{\textnormal{cav}}} \gg 1$, where $\Gamma$ is the spontaneous emission rate of the atom. In addition cavities need to be efficiently coupled to each other, i.e. $J \geq \gamma_{\textnormal{cav}}$.

There are three main potential technologies for the experimental implementation of coupled cavity arrays: fiber coupled microtoroidal cavities \cite{toroidal}, arrays of defects in photonic crystals \cite{defects} and superconducting qubits coupled via microwave stripline resonators \cite{superconducting}. Microtoroidal cavities are routinely produced in large arrays and can have high Q-factors \cite{105}. The cavities are coupled via tapered optical fibers which are placed close to the surface of the cavities and whose evanescent fields overlap thus allowing for photons to hop between neighbouring sites. The photon tunneling rate can be controlled by adjusting the distance between cavity and fiber. Moreover, these cavities can be made to interact with atoms in the strong coupling regime and single-cavity cooperativity parameters of $C \sim 50$ have already been demonstrated in an experiment \cite{58}. Some of the challenges for the realization of the effective Hamiltonian given by Eq.(\ref{Ham1}) using toroidal microcavities is that all the cavities  of the array should be coupled and tuned into resonance with each other.

Another promising candidate for implementing arrays of coupled cavities are atoms coupled to photonic band gap defect nanocavities. 
So far, large arrays of coupled cavities have been produced \cite{99} where photon hopping has been observed. Atomic impurities can be created inside these nanocavities and due to the cavities' small volume the interaction between the atom and the cavity is in the strong coupling regime with very large coupling constants \cite{60} and single-cavity cooperativity parameters up to $C \sim 10$. Spontaneous emission rates can be made small; however, cavity decay rates of photonic band gap cavities remain a limiting factor for the implementation of the dynamics of Eq.(\ref{Ham1}).

Coupling of two cavity QED systems formed by a Cooper-pair box coupled to a superconducting stripline resonator has not been achieved in an experiment yet, however, these systems possess some advantages that make them a possible platform for the implementation of the Hamiltonian (\ref{Ham1}). Strong coupling between the Cooper-pair box and the stripline resonator has been observed and very large single cavity cooperativity parameters $C\sim 10^{4}$ have been achieved already \cite{62}. Moreover, two Cooper-pair boxes have been strongly coupled to the cavity mode \cite{112,111} and theoretical investigations suggest that their number can be increased up to ten. These experimental results, combined with efficient cavity-cavity coupling make the circuit QED system a promising platform for the realization of the proposed algorithm.

\begin{figure}[t]
\includegraphics[width=85mm]{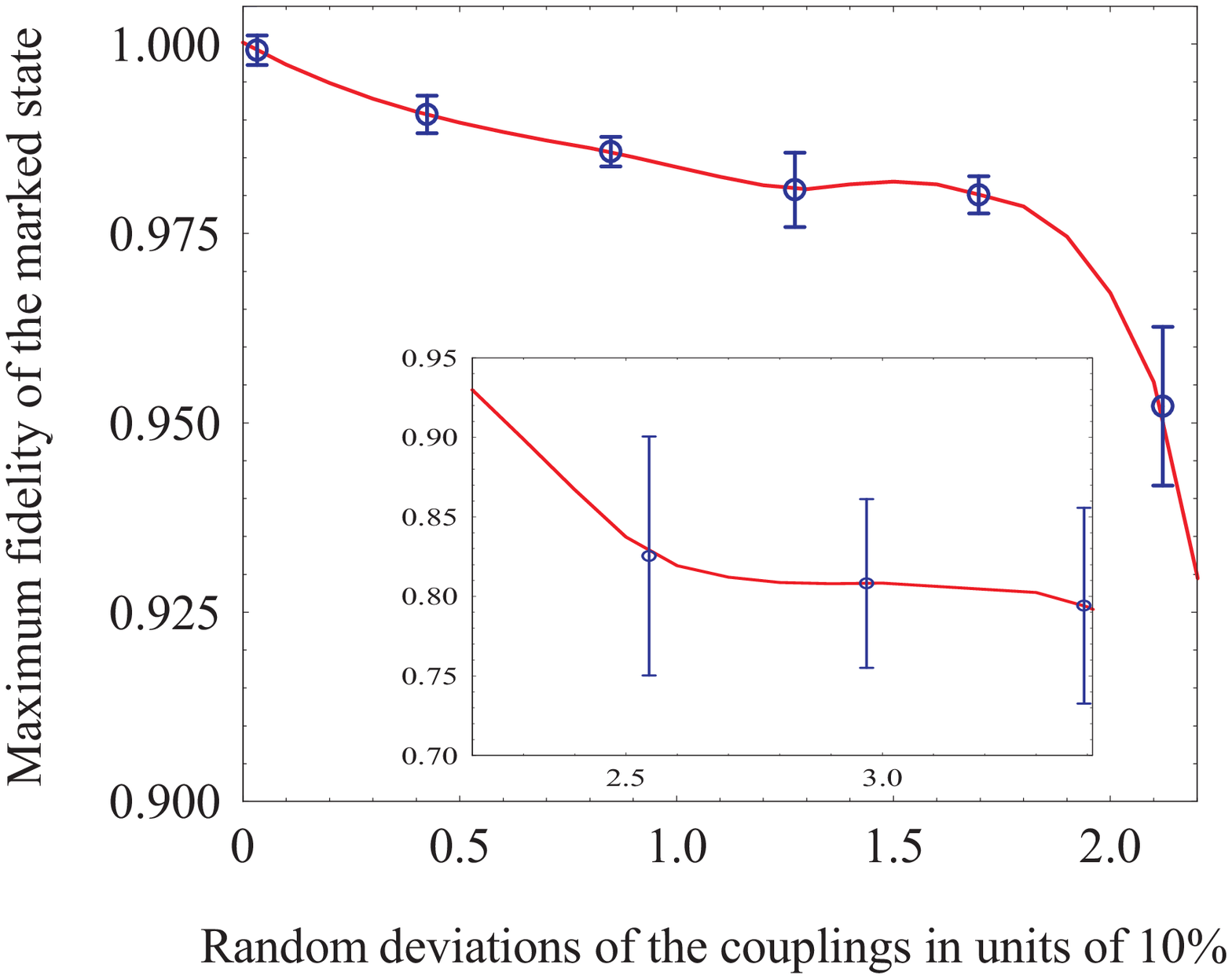}
\caption{Numerically calculated maximum fidelity of the marked state plotted against the random mean deviation of the different coupling constants $g_{i}^{\prime}$ in units of $10\%$.}
\label{fig3}
\end{figure}

In order to check the robustness of our protocol against possible experimental errors we have considered an array of 8 coupled microtoroidal cavities and have allowed for realistic inaccuracies in the resonant frequency of the cavities $\omega_{c}$. Another aspect which remains challenging and has the potential to reduce the fidelity of the protocol is the positioning of the atoms such that they all experience equal coupling constants. To estimate the influence of these errors we have performed a numerical simulation of the fidelity of the marked state as a function of a static deviation from the desired equal value of the effective coupling constants experienced by the individual qubits $g^{\prime}$. For the simulation shown in Fig. \ref{fig3} we have assumed a global pulse tuned on $\Delta^{\prime} = \delta $ with a hyperbolic-secant time dependence and a pulse area of 2$\pi$. Our calculations show that the marked element is successfully recovered with a probability of more than $70\%$ even when the deviations are of the order of $30\%$. This fidelity translates into success probability of about $20\%$ for a single step, which is comparable with the perfect classical search algorithm which has a maximum average success rate of $25\%$ for a single step. The error bars, however, increase rapidly when the mean fluctuations of the couplings are around $20\%$ but for smaller values they are of the order of few percents. This shows that we can achieve high fidelity with a relatively realistic system which can be crucial for the experimental implementation of the search algorithm.

In conclusion we showed how symmetric cavity-qubit couplings can be used to implement a robust and efficient quantum search algorithm in a linear chain of $N$ coupled cavities doped with atoms whose hyperfine states are used to encode the logical qubits. The algorithm is implemented by employing global laser fields in such a way that the number of the required physical steps become equal to the number of logical steps, i.e., $\sim \mathcal{O}(\sqrt{N})$. This is an advantage compared to other implementations which in addition to these query steps require many more physical non-query steps. We achieve the speed up with regard to physical steps by using the natural evolution of the system, which already implements the global reflection operation required by the algorithm. We also allow for realistic deviations from the symmetric cavity-qubit couplings that can arise in an experimental setup and find that our proposal is quite robust against such imperfections.

\acknowledgments
We would like to acknowledge useful discussions with S. Ivanov and N. Vitanov. Financial support was provided by the National Research Foundation and the Ministry of Education, Singapore.


\end{document}